# Competing Phases in Epitaxial Vanadium Dioxide at Nanoscale

Yogesh Sharma[1], Martin V. Holt[2], Nouamane Laanait[3], Xiang Gao[1], Ilia Ivanov[3], Liam Collins[3], Changhee Sohn[1], Zhaoliang Liao[1], Elizabeth Skoropata[1], Sergei V. Kalinin[3], Nina Balke[3], Gyula Eres[1], Thomas Z. Ward[1], and Ho Nyung Lee[1*]

[1]Materials Science and Technology Division, Oak Ridge National Laboratory, Oak Ridge, TN 37831, USA

[2]Center for Nanoscale Materials, Argonne National Laboratory, Lemont, IL 60439, USA

[3]Center for Nanophase Materials Sciences, Oak Ridge National Laboratory, Oak Ridge, TN 37831, USA

*hnlee@ornl.gov

**ABSTRACT**

Phase competition in correlated oxides offers tantalizing opportunities as many intriguing physical phenomena occur near the phase transitions. Owing to a sharp metal-insulator transition (MIT) near room temperature, correlated vanadium dioxide ($VO_2$) exhibits a strong competition between insulating and metallic phases that is important for practical applications. However, the phase boundary undergoes strong modification when strain is involved, yielding complex phase transitions. Here, we report the emergence of the nanoscale M2 phase domains in $VO_2$ epitaxial films under anisotropic strain relaxation. The phase states of the films are imaged by multi-length-scale probes, detecting the structural and electrical properties in individual local domains. Competing evolution of the M1 and M2 phases indicates a critical role of lattice-strain on both the stability of the M2 Mott phase and the energetics of the MIT in $VO_2$ films. This study demonstrates how strain engineering can be utilized to design phase states, which allow deliberate control of MIT behavior at the nanoscale in epitaxial $VO_2$ films.

In strongly correlated materials, metastability and phase competition play an important role in determining the phase transition pathway.[1–5] The metal-insulator transition (MIT) in strongly correlated vanadium dioxide (VO$_2$) is believed to be triggered by the collaborative Mott-correlation and Peierls-distortion mechanisms, which is still under debate due to the observation of metastable insulating phases across the MIT.[6–17] While tremendous efforts have been devoted to understanding the origin of the phase transition and related electronic behavior, identifying intermediate phases of distinct electronic and structural compositions across the MIT has also been one of the main subjects in VO$_2$ research.[13,14,18–27]

Bulk VO$_2$ is known to undergo a phase transition from an insulating monoclinic (M1) phase to a metallic rutile (R) phase close to room temperature ($T_C$ ~340 K in bulk).[6,7] Recently, strain control has enabled the observation of a more complex phase transition,[11,15,28–41] where another insulating monoclinic (M2) phase can be found as an intermediate transitional phase between the M1 and the R phases.[19,22,23,28,42,43] Tensile strain is found to induce a M1→M2→R transition pathway, whereas compressive strain shows a M1→R transition depending on the degree and sign of the applied strain field.[22,28,44,45] The M2 phase is believed to have comparable free energy to M1 in bulk VO$_2$, which can be stabilized by modulating strain.[22] This is important since stabilizing the M2 phase may find use in Mottronics and novel electronic transport applications.[46,47] The metastability of the M2 phase also highlights its importance as a Mott insulator, which could potentially settle the debate about the coupled nature of Peierls and Mott-Hubbard mechanisms responsible for the MIT in VO$_2$.[48,49] Combined local imaging of structural and electronic properties of the M2 phase across the transition are lacking. This is part of the reason why it is so difficult to understand the underlying mechanism that drives this step in the nanoscale phase transition process in VO$_2$.

Here, we investigate the local structural and electronic properties of differently strained VO$_2$ epitaxial films to identify such intriguing phases. Using multimodal imaging techniques, including large area probes averaging over multiple phase domains of different character, we directly image the coexistence of distinct metallic and insulating nanoscale phase domains across the MIT. We observe that the M2 phase emerges in forms of nanophase domains at the onset of strain relaxation, which is attributed to the complex localized stress distribution in the film. Understanding the origin of these mesoscopic structural and electronic inhomogeneities would greatly benefit the nanoscale engineering of correlated phases in VO$_2$ films.

VO$_2$ epitaxial films were deposited using pulsed laser epitaxy. A bulk ceramic VO$_2$ target of M1 phase was ablated by a KrF excimer laser (248 nm in wavelength) at a laser fluence of 1.2 Jcm$^{-2}$ and at a laser repetition rate of 5 Hz. The growth temperature and oxygen partial pressure were optimized to 300 °C and 12 mTorr, respectively. To avoid oxygen deficiencies, the samples were cooled to room temperature right after deposition at higher oxygen pressure of 100 Torr.[50] X-ray measurements were performed by a four-circle high-resolution X-ray diffractometer (X'Pert Pro, Panalytical) using the Cu-K$\alpha$1 radiation. A standard four-point probe method was used to perform the electrical transport measurements of VO$_2$ films in a temperature range varied from 200 to 398 K by a physical property measurement system (Quantum Design). SPM studies were performed with a commercial AFM system (Cypher from Asylum Research an Oxford Instruments Company) equipped with a variable-bandwidth current amplifier (FEMTO DLPCA-200). The heating-cooling stage (Integrated temperature control from Asylum research) was used for temperature dependent c-AFM measurements. A conducting tip; PtSi-coated tip (PtSi-FM-20, Nanosensors) was used for c-AFM measurements. c-AFM scans were recorded by applying a bias of 0.4 V to the tip, while sample was grounded.

Nanoscale scanning probe X-ray-diffraction measurements were performed at the Hard X-ray Nanoprobe (HXN) beamline operated by the Center for Nanoscale Materials in partnership with the Advanced Photon Source at Argonne National Laboratory. The HXN beamline uses Fresnel zone plate optics with 20 nm outermost zone width to provide an X-ray beam spot with a 20 nm FWHM lateral resolution and a position stability of 2-5 nm relative to the sample area.[51,52] The scanning probe diffraction microscopy measurements were carried out at an X-ray energy of 9 keV in reflection geometry with the outgoing diffraction patterns collected by a MediPix3 Pixel Array Detector with 516x516 µm$^2$ pixels positioned 800 mm from the sample. X-ray diffraction Microscopy (XDM) was performed with the XRIM instrument at sector 33-ID of the Advanced Photon Source at Argonne National Laboratory. The imaging was carried at an X-ray energy of 10 keV in a Bragg diffraction geometry, using an illumination of ~ 12x12 µm$^2$ achieved with a pair of dynamically bent Kirkpatrick-Baez mirrors. A phase Fresnel Zone plate (100 µm diameter, 60 nm outermost zone width) was used as objective lens, leading to an effective pixel size at the sample location of 15 nm. Micro-Raman measurements were performed using a Renishaw 1000 confocal Raman microscope. A temperature control microscopy state (Linkam Scientific) stage was used for temperature dependent Raman studies. The 532 nm diode-pumped solid-state laser (Cobolt) with a laser power 10 mW was used to excitation. Raman spectra were collected in back scattering geometry using 100× objective (NA= 0.75). Raman spectrometer was equipped with 1800 lines/mm grating which was centered at 500 cm$^{-1}$. For Raman mapping, we collected 4118 spectra (10s integration time, and using 500 nm step size) on a 25×22 µm$^2$ region of the sample.

Epitaxial VO$_2$ films with thicknesses of 12, 28, and 48 nm were grown on (001) TiO$_2$ and Nb-doped TiO$_2$ (Nb:TiO$_2$) single crystal substrates using pulsed laser epitaxy. X-ray diffraction is used to confirm film thicknesses and crystalline quality [Fig. 1(a)]. X-ray reciprocal space maps (RSMs) around the TiO$_2$ 202 and 022 Bragg peaks show a gradual strain relaxation with increasing film thickness. We observe that the lattice strain is fully maintained in the 12 nm thick film and relaxed in 28 nm and 48 nm films (see the supplementary material). The MIT characteristics of VO$_2$ films were investigated by *dc* transport measurements as a function of temperature across the phase transition. The temperature dependent resistivity curves of VO$_2$ films are shown in Fig. 1(b). In this film thickness range, the MIT temperature ($T_{MIT}$) was observed to span the range of 296-328 (±5) K with a hysteresis value of 8-14 (±2) K. The higher resistivity was observed for the thicker films compared to the coherently strained 12 nm VO$_2$ film, which is consistent with previous reports.[35,36] Interestingly, for the thicker films, multiple steps were observed in resistivity vs temperature curves [Fig. 1(b)] due to the coexistence of M1 and M2 phases and their transition to rutile-R phase across the MIT. These trends on the transport properties were previously believed to be due to the emergence of either an intermediate insulating phase or anisotropic strain relaxation.[10,29,30,53] The origin of multistep MIT behavior in thicker films is also speculated to be due to the formation of microcracks [36], as strain relaxes in tensile strained VO$_2$ films (see the supplementary material for more details). The unusual strain relaxation can significantly affect the crystal lattice symmetry and electronic states. The resulting changes in the displacement (tilting) and pairing of vanadium atoms can then act to stabilize different VO$_2$ phases [Fig. 2(a)].[13,15,19,22]

To gain insight into the local electronic properties, we employed temperature dependent conducting atomic force microscopy (c-AFM) to study the nanoscale phase evolution in VO$_2$ films with strain. c-AFM images across the MIT temperature together with the surface topography images at 298 K are shown in Figs. 2(b-d). These images show significant changes in both surface and local conductivity of the films with strain relaxation. Below the MIT temperature on 12 nm films, we observe nearly homogeneous low current (insulating state) in the c-AFM image at 278

K. The resistance gradually decreases to a very high current value (metallic state)—indicating that a fully strained film undergoes a direct transition from M1 to R phase.[32] As the film thickness is increased to 28 nm and 48 nm in Figs. 2(b-d), we observe the presence of stripe domains throughout the film having different electrical conductivities. Such domains of different electronic states were previously observed in strain relaxed $VO_2/TiO_2$ (001) films using infrared (IR) nanoimaging techniques.[12,29,30,53] However, in contrast to previous reports in which the observations were limited to the qualitative information from the IR scattering amplitude and phase images, our results provide details on the electronic phase evolution by measuring the current of a single localized domain at nanoscale (see the supplementary material). The c-AFM results are consistent with the structural transition pathway of M1→M1+M2 →R across the films' MITs. These results indicate that the formation of domains of different electronic states most likely originates from the strain relaxation-induced structural and electronic phase inhomogeneities in $VO_2$ films. Direct confirmation of this distinct structural phase transition pathway is challenging due to small feature size.

To study the local structure of these domains, we performed synchrotron-based nanoscale XRD and scanning probe diffraction microscopy measurements on 48 nm sample.[54] The measurements were carried out using the source-detector-sample configuration shown in Fig. 3(a), where the incident X-rays were focused by a Fresnel-zone-plate to a 20 nm FWHM focused spot and the scattered X-rays were collected by a *2D* pixel array detector.[55] The inset of Fig. 3(a) shows an AFM topography image of the sample across the area of detection. The room temperature XRD pattern is shown in Fig. 3(b). The Bragg peak at *2θ = 29.2°* originates from the 002 reflection of the M1 phase, while the Bragg peak centered at *2θ = 28.90°* corresponds to the monoclinic M2 phase. To spatially map the distribution of M1 and M2 phases in the film, nanoscale scanning probe diffraction microscopy measurements were conducted while setting the detector at a specific scattering angle, $2\theta$, corresponding to each peak, and recording the intensity as a function of position at room temperature. Figures 3(c) and 3(d) show two-dimensional 5 μm × 5 μm spatial maps of the X-ray intensity with the detector set at the scattering angle $2\theta = 28.90°$ and $29.58°$, respectively. Interestingly, we observe a higher intensity of the M2 phase along the boundaries to the cracks as well as for the domains of narrower width. These observations are consistent with expectations from AFM results, where the c-axis elongated lattice strain locally stabilizing the higher lattice constant M2 phase at the expense of the M1 phase near the boundary regions. Temperature-dependent nanoscale XRD results further demonstrate that there is almost no X-ray intensity corresponding to the cracks and that the M1/M2 phase domains convert to R phase at higher temperatures above the MIT (see the supplementary material). The nanoscale XRD results strongly agree with the c-AFM findings that the M2 phase together with the M1 phase can be revealed at room temperature in strain-relaxed $VO_2$ film.

To obtain further proof that the two observed $VO_2$ regions are distinct crystallographic phases, we carried out additional measurements using synchrotron full-field X-ray Diffraction Microscopy (XDM).[56] Like scanning probe diffraction microscopy (Fig. 3), the imaging contrast in large area XDM-probe is entirely dependent on Bragg diffraction.[57] XDM images were acquired using the $\bar{1}12$ and 002 film reflections (referenced against the rutile $TiO_2$) with orthogonal *HK* Miller indices (Fig. 4). Peak splitting in both the *HHL* and *00L* reflections is a characteristic signature of M1 and M2 phases [Figs. 4(a) and 4(b)]. XDM images were acquired at the Bragg reflections with a reciprocal lattice vector tuned to match the M1 (M2) peak. This allowed a selective probe of the distribution of M1 (M2) phase in real-space with a lateral resolution of roughly 100 nm. Stable intermediate M2 phase domains are observed and are mostly localized along the boundary regions.

Raman probe is extensively used to identify the different structural polymorphs of $VO_2$ at the microscale in pure and doped bulk single crystals of $VO_2$.[20,21,45,58] We employ micro-Raman spectroscopy and mapping measurements to understand the phase transition and evolution of the M2 phase with temperature. Figure 5(a) shows the room temperature Raman spectra of a $TiO_2$ substrate and the $VO_2/TiO_2$ films. The Raman spectrum of a metallic 12 nm thick $VO_2$ film is mainly dominated by the Raman signals from the $TiO_2$ substrate, whereas the spectra of the 28 nm and 48 nm films show the peaks at 194, 224, 310, 352, 382, and 495 $cm^{-1}$, which all corresponds to a M1 symmetry.[20,31,42,43] Interestingly, additional peaks at 235 $cm^{-1}$ (split from the 224 $cm^{-1}$ mode) and 434 $cm^{-1}$ of the M2 symmetry[45] can be observed in Raman spectra of the 28 nm and 48 nm films.[20,42] The splitting of the 224 $cm^{-1}$ mode is considered to be a distinct signature for the presence of the M2 phase in bulk $VO_2$ [59] and helps confirm the distribution of M2 phase over the large area of the samples. Figures 5(b) and (c) show the Raman intensity maps of the 235 $cm^{-1}$ and 434 $cm^{-1}$ peaks of the M2 phase for an area of 25 µm × 22 µm on the 48 nm film. Intensity map details clearly show regions of similar shape and scale to the domains observed in c-AFM and X-ray microscopy results. Figure 5(d) shows the Raman spectra recorded as temperature is lowered across the phase transition—from metallic at 380 K to insulating at 200 K. The M2 phase peak at 235 $cm^{-1}$ starts to evolve at around 240 K [Fig. 5(e)] and vanishes near 320 K, which is consistent with the transition being tied to the MIT. The phase transition is observed to be perfectly reversible as confirmed by recording the Raman spectra while heating the sample above the MIT temperature (see the supplementary material). Raman findings confirm the presence of the M2 phase in our strain relaxed epitaxial VO$_2$ films and support the claim of c-AFM measurements regarding the structure transition pathway of M1→M1+M2 →R across the MIT.

In summary, observation of the emergence of stable intermediate phase states in epitaxial VO$_2$ films have been performed by multimodal imaging techniques at various length scales. The critical role of lattice strain on the stability of the M2 Mott phase and the energetics of the metal-insulator transition (MIT) in epitaxial VO$_2$ films is demonstrated. An unusual local strain-dependent relaxation process allows the simultaneous creation of multiscale domains of M1, M2, and R phases. Our findings provide an insight into the complicated mesoscopic electronic and structural inhomogeneities across VO$_2$'s MIT and may lead to the engineering of these correlated phases for electronic and photonic nanodevice applications.

**Supplementary Material:** See supplementary material for X-ray reciprocal space mapping (RSM) measurements, scanning transmission electron microscopy (STEM), the local I-V curves in CAFM mode, scanning probe X-ray diffraction at higher temperatures, and micro-Raman spectroscopy and imaging.

**ACKNOWLEDGMENTS**

This work was supported by the U.S. Department of Energy (DOE), Office of Science, Basic Energy Sciences (BES), Materials Sciences and Engineering Division (synthesis, microstructural characterization, and scanning probes) and as part of the Computational Materials Sciences Program (structural characterization). Raman measurements were performed as user projects at the Center for Nanophase Materials Sciences, which is sponsored at Oak Ridge National Laboratory by the Scientific User Facilities Division, BES, U.S. DOE. Use of the Center for Nanoscale Materials and the Advanced Photon Source, both Office of Science user facilities, was supported by the U.S. Department of Energy, Office of Science, Office of Basic Energy Sciences, under Contract No. DE-AC02-06CH11357.**REFERENCES**

[1] M. Imada, A. Fujimori, and Y. Tokura, Rev. Mod. Phys. **70**, 225 (1998).

[2] K. H. Ahn, T. Lookman, and A.R. Bishop, Nature **428**, 401 (2004).

[3] M. Fäth, Science **285**, 1540 (1999).

[4] M. Uehara, S. Mori, C.H. Chen, and S.-W. Cheong, Nature **399**, 560 (1999).

[5] A. Moreo, Science **283**, 2034 (1999).

[6] F. J. Morin, Phys. Rev. Lett. **3**, 34 (1959).

[7] N. F. Mott, *Rev. Mod. Phys.* **40**, 677–683 (1968).

[8] J. B. Goodenough, Journal of Solid State Chemistry **3**, 490 (1971).

[9] R. M. Wentzcovitch, W. W. Schulz, and P. B. Allen, Physical Review Letters **72**, 3389 (1994).

[10] S. Kumar, J. P. Strachan, M. D. Pickett, A. Bratkovsky, Y. Nishi, and R. S. Williams, Advanced Materials **26**, 7505 (2014).

[11] D. Lee, B. Chung, Y. Shi, G.-Y. Kim, N. Campbell, F. Xue, K. Song, S.-Y. Choi, J. P. Podkaminer, T. H. Kim, P. J. Ryan, J.-W. Kim, T. R. Paudel, J.-H. Kang, J. W. Spinuzzi, D. A. Tenne, E. Y. Tsymbal, M. S. Rzchowski, L. Q. Chen, J. Lee, and C. B. Eom, Science **362**, 1037 (2018).

[12] X. Chen, D. Hu, R. Mescall, G. You, D. N. Basov, Q. Dai, and M. Liu, Advanced Materials **0**, 1804774 (2019).

[13] J. Cao, E. Ertekin, V. Srinivasan, W. Fan, S. Huang, H. Zheng, J. W. L. Yim, D. R. Khanal, D. F. Ogletree, J. C. Grossman, and J. Wu, Nature Nanotechnology **4**, 732 (2009).10

**Figure 1**

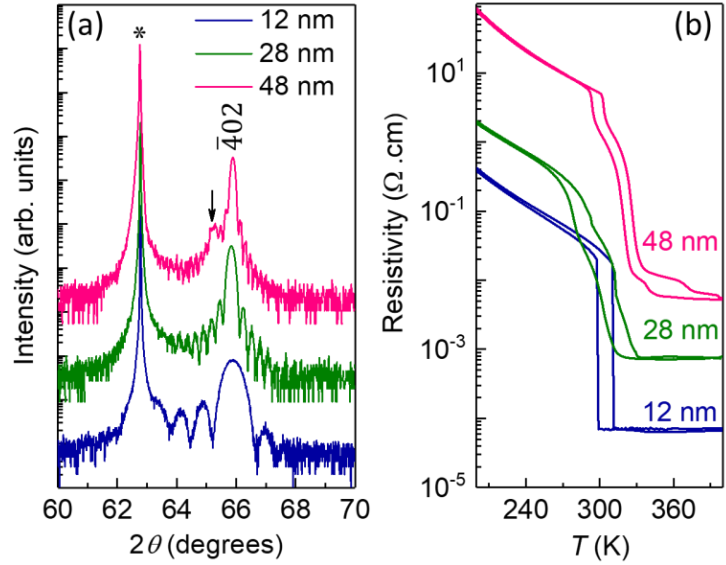

**FIG. 1.** Structure and transport properties of differently strained epitaxial VO$_2$ films. (a) Room temperature XRD θ–2θ line scans of VO$_2$ films grown on (001) TiO$_2$ substrates with varying the film thickness ranging from 12 to 48 nm. The peak from 002 reflection of the TiO$_2$ substrate is labeled as an asterisk (*), and an additional peak close to the monoclinic $\bar{4}02$ peak in a 48 nm film is labeled as an arrow (↓) sign. (b) Resistivity vs. temperature curve of VO$_2$ thin films where overall resistivity increases with increasing film thickness. The observation of an additional peak in XRD and multistep MIT behavior indicate the presence of stable intermediate insulating (M2) phase in fully relaxed 48 nm films.

# Figure 2

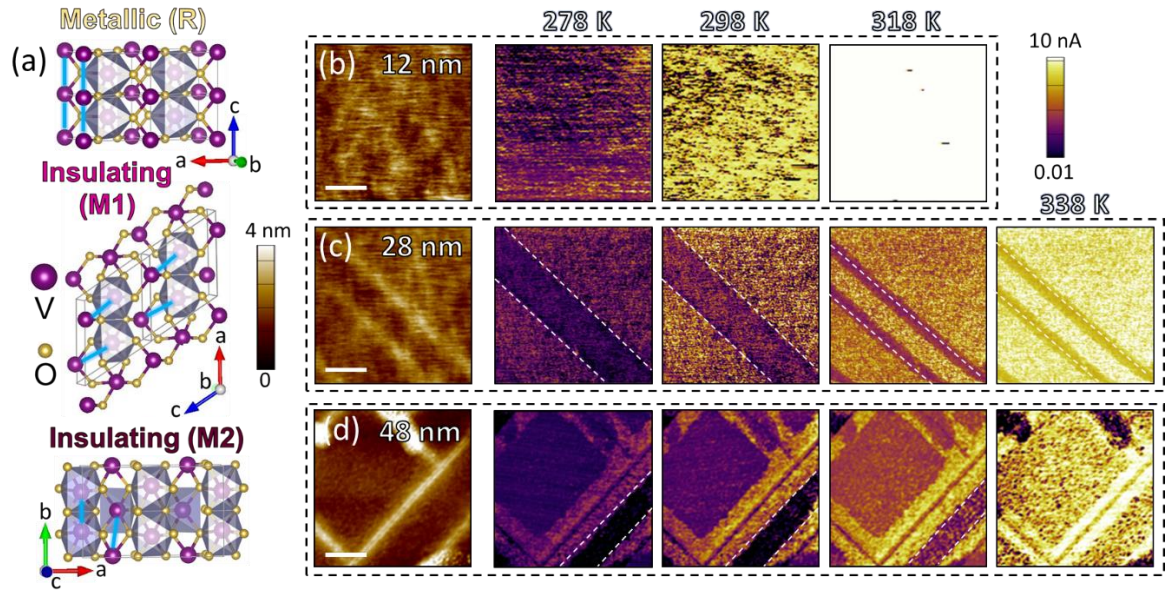

**FIG. 2.** Strain-relaxation induced changes in morphological and local-conducting properties across MIT. (a) Schematics of crystal structures of different electronic phases of $VO_2$; V-atoms' chains are periodic and linearly arrayed in R phase, whereas all the chains are tilted and dimerized along $c_R$ axis in the M1 phase. M2 is different from M1 in that only half of the chains are dimerized and other half of the chains remain straight (undistorted) like in R phase. The surface topography (at 298 K, second column from left) and c-AFM images of (b)12 nm, (c) 28 nm, and (d) 48 nm films at various temperatures across the MIT in Fig. 1(b). The formation and temperature dependent evolution of insulating (M2) phase nanodomains (marked by white dashed lines) can be observed in strain relaxed 28 and 48 nm thick films, whereas in fully strained 12 nm film the local electronic phase transition is continuous with nearly homogenous single insulating/metallic domain. Scale bar is 600 nm.

**Figure 3**

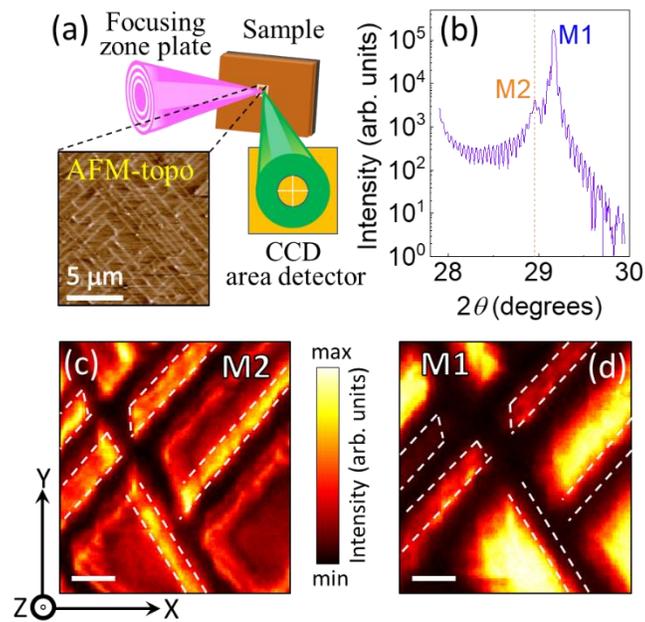

**FIG. 3.** Nanoscale observation of M1 and M2 phase domains at room temperature. (a) Schematic of nanoscale X-ray diffraction setup where the incident X-rays (9 keV) are focused by a Fresnel Zone plate to a 20 nm spot on the sample, while the exiting X-rays from the sample are collected by a CCD area detector. Inset of (a) shows an AFM topographic image of the 48 nm film across the area of detection. (b) Room temperature nano X-ray diffraction pattern represent the Bragg intensity from the monoclinic M2 phase (dashed line) and 002 reflection of M1 phase of $VO_2$. Two dimensional spatial maps of the 5 μm×5 μm area showing the variation in the intensity of the Bragg peak from (c) M2 (white dashed lines) and (d) M1 phases of $VO_2$ obtained with the detector angle set at corresponding (2θ) peak positions. Scale bar in (c) and (d) is 1μm.

# Figure 4

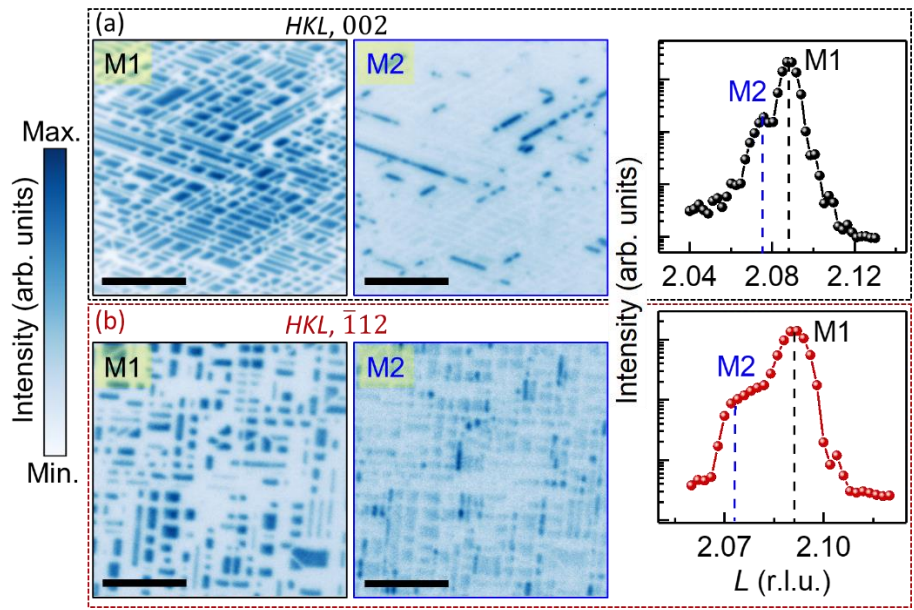

**FIG. 4.** Large area probes average over multiple phase domains of M1/M2. (a) Full-field X-ray diffraction microscopy (XDM) images taken at the 002 Bragg reflection for each M1/M2 phase indicate the presence of M2 phase, which is largely localized along boundaries with M1. (b) XDM images taken at the off-specular $\bar{1}12$ reflection provides direct structural evidence for identification of the M1 and M2 domains with distinct crystallographic phases. Some thickness variations within the same phase are also observed, given that the intensity in each XDM image is proportional to the square of the number of diffracting ($\bar{1}12$) or (002) planes. The scale bar is 3 µm.

**Figure 5**

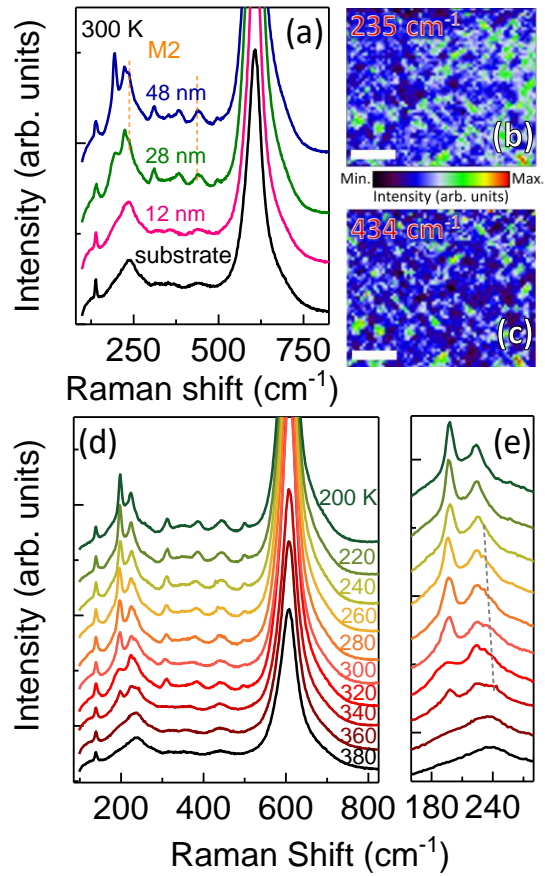

**FIG. 5.** Micro-Raman mapping of M2 phase domains and temperature-driven structural phase evolution. (a) Room temperature Raman spectra of $VO_2$ films (12, 28, 48 nm in thickness), including $TiO_2$ substrate; the spectrum of a 12 nm thick $VO_2$ film is dominated by the Raman signals from the $TiO_2$ substrate whereas the spectra of 28 and 48 nm thick films corresponds to the monoclinic M1 symmetry with two additional characteristic Raman modes from M2 symmetry at 235 and 434 $cm^{-1}$ (dashed lines). Raman intensity mapping of (b) 235 $cm^{-1}$ and (c) 434 $cm^{-1}$ peaks of M2 phase in a 48 nm film at room temperature. (d) and (e) represent the temperature-dependent Raman spectra of 48 nm $VO_2$ film. The temperature evolution of M2 phase (dashed line in (e)) supports the c-AFM observations regarding the structural-transition pathway of M1→M1+M2 →R across the MIT. Scale bar in (b) and (c) is 6 µm.